\PassOptionsToPackage{utf8}{inputenc}
\documentclass{bioinfo}

\usepackage{comment}
\usepackage{graphicx}
\usepackage{amsmath}
\usepackage{multirow}
\usepackage{comment}

\usepackage{float}
\floatstyle{plaintop}
\restylefloat{table}

\usepackage[table,xcdraw]{xcolor}

\usepackage{xcolor}
\usepackage[colorlinks = true,
            linkcolor = blue,
            urlcolor  = blue,
            citecolor = blue,
            anchorcolor = blue]{hyperref}

\newcommand{\MYhref}[3][blue]{\href{#2}{\color{#1}{#3}}}%

\copyrightyear{2015} \pubyear{2015}

\access{Advance Access Publication Date: Day Month Year}
\appnotes{Manuscript Category}

\DeclareUnicodeCharacter{2212}{-}
\begin{document}
\firstpage{1}

\subtitle{Sequence Analysis}

\title[iPromoter-BnCNN]{iPromoter-BnCNN: a Novel Branched CNN Based Predictor for Identifying and Classifying Sigma Promoters}
\author[R.Amin \textit{et~al}.]{Ruhul Amin\,$^{\text{\sfb 1,}}$, Chowdhury Rafeed Rahman\,$^{\text{\sfb 1}*}$, Md. Habibur Rahman Sifat\,$^{\text{\sfb 1}}$, Md Nazmul Khan Liton\,$^{\text{\sfb 1}}$, Md. Moshiur Rahman\,$^{\text{\sfb 1}}$, Sajid Ahmed\,$^{\text{\sfb 1,}}$ and Swakkhar Shatabda\,$^{\text{\sfb 1,}*}$} 
\address{$^{\text{\sf 1}}$Department of Computer Science and Engineering, United International University, Dhaka, 1207, Bangladesh\\}

\corresp{$^\ast$Swakkhar Shatabda}

\history{Received on XXXXX; revised on XXXXX; accepted on XXXXX}

\editor{Associate Editor: XXXXXXX}

\abstract{\textbf{Motivation:}  Promoter is a short region of DNA which is responsible for initiating transcription of specific genes. Development of computational tools for automatic identification of promoters is in high demand.  According to the difference of functions, promoters can be of different types. Promoters may have both intra and inter class variation and similarity in terms of consensus sequences. Accurate classification of various types of sigma promoters still remains a challenge. \\
\textbf{Results:} We present \textbf{iPromoter-BnCNN} for identification and accurate
classification of six types of promoters - $\sigma^{24}, \sigma^{28}, \sigma^{32}, \sigma^{38}, \sigma^{54}, \sigma^{70}$. It is a Convolutional Neural Network (CNN) based classifier which combines local features related to monomer nucleotide sequence, trimer nucleotide sequence, dimer structural properties and trimer structural properties through the use of parallel branching. We conducted experiments on a benchmark dataset and compared with two state-of-the-art tools to show our supremacy on 5-fold cross-validation. Moreover, we tested our classifier on an independent test dataset. \\
\textbf{Availability:}  Our proposed tool iPromoter-BnCNN web server is freely available at \MYhref{http://103.109.52.8/iPromoter-BnCNN}{http://103.109.52.8/iPromoter-BnCNN}. The runnable source code can be found \MYhref{https://colab.research.google.com/drive/1yWWh7BXhsm8U4PODgPqlQRy23QGjF2DZ}{here}.
\textbf{Contact:} \href{name@bio.com}{rafeed@cse.uiu.ac.bd}\\
\textbf{Supplementary information:} Supplementary data (benchmark dataset, independent test dataset, structural property information and model .h5 files) are available at \textit{Bioinformatics}
online.}

\maketitle

\section{Introduction}
Promoters are small regions near gene containing 100 to 1000 base-pairs. For transcription occurrence, RNA polymerase must bind near the promoter. Bacteria with prokaryotic cell type has promoters consisting of a purine at the transcription start site (TSS). It contains specific hexamers centered at -10 and -35 (\cite{busby1994promoter}, \cite{feng2017irna}). There are several sigma factors in the RNA polymerase of Escherichia coli bacteria, which are dependent on environment and gene. As a result, sigma factors are used as distinguishing elements of promoter sequences found in DNA. Each of the six different types of sigma factors such as $\sigma^{24}, \sigma^{28}, \sigma^{32}, \sigma^{38}, \sigma^{54}, \sigma^{70}$ has different functions. For example, $\sigma^{70}$ factor is responsible for transcription of most of the genes under normal condition (\cite{gruber2003multiple}). On the other hand, $\sigma^{24}$ factor is responsible for heat shock response (\cite{raina1995rpoe}). Similarly, $\sigma^{28}, \sigma^{32}, \sigma^{38}$ and $\sigma^{54}$ are responsible for flagellar genes, heat shock response,  stress response during the transition from exponential growth phase to the stationary phase of E. coli (\cite{jishage1995regulation}) and nitrogen metabolism, respectively (\cite{janga2007structure}).

Molecular techniques for promoter identification or classification is costly in terms of time and money which is why computational methods are more popular (\cite{towsey2008cross}). Promoters normally differ from the consensus at one or more positions. So, it is challenging to precisely predict promoters through traditional methodology.

Recently, a few computational methods have been proposed to classify DNA sequences as promoters or non-promoters, some aiming at identifying a certain class of sigma promoters. For instance, \cite{coelho2018bacillus} provided BacSVM+, a software package using LibSVM library for promoter prediction in Bacillus subtilis. Work of \cite{e2014dna} integrated DNA duplex stability as feature of neural network to identify $\sigma^{28}$ and $\sigma^{54}$ class of promoter in E. coli bacteria. \cite{lin2014ipro54} developed iPro54-PseKNC which performs the same task using SVM classifier based on pseudo k-tuple nucleotide composition (PseKNC). \cite{li2015deep} applied a deep feature selection (DFS) model on enhancer-promoter classification. \cite{lin2017identifying} used pseudo nucleotide composition for feature extraction in order to identify $\sigma^{70}$ promoters in prokaryotes using SVM. \cite{he201870propred} used PSTNPSS(Position-specific trinucleotide propensity 
based on single-stranded characteristic) and PseEIIP(Electron-ion potential values for trinucleotides) features while \cite{rahman2019ipro70} used multiple windowing and minimal features for the same task. \cite{rahman2019ipromoter} developed iPromoter-FSEn for performing the same task using feature subspace based ensemble classifier achieving an impressive accuracy of 86.32\%. \cite{umarov2017recognition} trained CNN based architecture on the same promoter type in E. coli. \cite{liu2017ipromoter} developed iPromoter-2L which can identify promoter and can classify them into six types. They used random forest PseKNC. \cite{zhang2019multiply} proposed MULTiPly for the same task using both local (k-tuple nucleotide composition, dinucleotide based auto covariance) and global information (bi-profile Bayes and KNN feature encodings). They applied F-score feature selection method to identify feature from each category giving the best prediction results.

\cite{shahmuradov2017btssfinder} worked on predicting transcription start sites (TSSs) in five types of E. coli sigma promoters such as - $\sigma^{24}, \sigma^{28}, \sigma^{32}, \sigma^{38}$ and $\sigma^{70}$ though they did not work on E. coli sigma promoter classification. Only
\cite{liu2017ipromoter} and \cite{zhang2019multiply} proposed computational methods (iPromoter-2L and MULTiPly) for classifying sigma promoters into six classes in E. coli bacteria. The sensitivity and specificity of promoter classification showed opposing behavior for iPromoter-2L. For example, for $\sigma^{28}, \sigma^{32}, \sigma^{38}$ and $\sigma^{54}$, iPromoter-2L showed specificity of higher than 99\%, but the sensitivity was lower than 54\%. The promoter classification performances of the binary sub-classifiers used in MULTiPly were impressive. For example, the first sub-classifier showed 85.24\% accuracy in $\sigma^{70}$ promoter type identification. The sensitivity and specificity was 87.27\% and 86.57\%, respectively. We follow the stage by stage binary classification method used in MULTiPly in this work. The main limitation of MULTiPly was the selection of the basic features to work with. Different combination of different heterogeneous features led to different prediction results. Effective selection of basic and essential features for the classification model is a difficult problem to solve. Through trial and error, the authors selected features that achieved satisfactory prediction performance.  

We propose iPromoter-BnCNN, a one dimensional CNN based classifier which can identify sigma promoter and can classify sigma promoter into the six specified classes in E. coli bacteria. Four parallel branches of one dimensional convolution filters learn and and extract important local features related to monomer, trimer nucleotide sequence and dimer, trimer structural properties simultaneously. Dense layers at the end of our designed model combine these extracted features and perform the classification task. We use the same model architecture for all of our binary classifiers. From the training samples, each classifier learns weights and importance of features automatically. We compare our method with state-of-the-art tools for E. coli sigma promoter identification and classification and show the effectiveness of our method.    

\section{ Materials and Methods}
\begin{figure}[h]
    \centering
    \includegraphics[width=1.0\columnwidth]{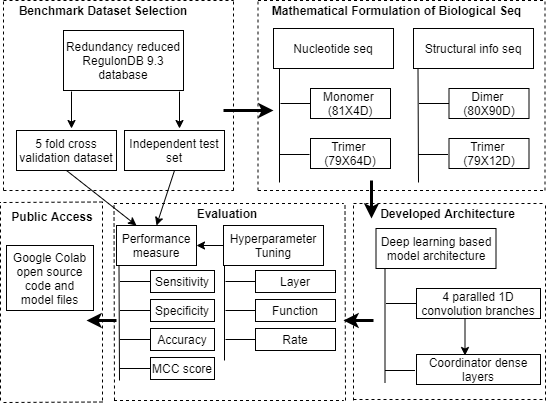}
    \caption{System Overview of iPromoter-BnCNN}
\label{fig:overview}
\end{figure}

We followed Chou’s five-step rules (\cite{chou2011some}) for more effective presentation of our research work. A series of recent publications (\cite{xu2013isno}, \cite{xu2014ihyd}, \cite{liu2015identification}, \cite{jia2016isuc}, \cite{chen2016irna}, \cite{feng2017irna}, \cite{liu2017ipromoter}, \cite{zhang2019multiply}) comply with this standard. Briefly, the five steps are: (i) valid benchmark dataset selection (ii) biological sequence sample formulation with mathematical expression (iii) powerful algorithm introduction for prediction purpose (iv) predictor performance evaluation using cross validation (v) public access establishment to the constructed predictor. Our system overview has been provided in Figure \ref{fig:overview} in light of the five steps described. We describe each of these steps in detail in the following subsections.  

\subsection{Benchmark Dataset}
One benchmark dataset is good enough to prove the effectiveness of a certain method when K-fold cross-validation is used, because such evaluation takes into account the results obtained from K number of disjoint training and validation sets (\cite{chou2007recent}). We have used the same E. coli bacteria promoter dataset as of \cite{liu2017ipromoter} and \cite{zhang2019multiply} for comparison purpose in terms of sigma promoter identification and classification into sub types. All promoter samples of the used dataset are experimentally verified (each has 81 bp). They have been collected from the RegulonDB database (Version 9.3) (\cite{gama2016regulondb}). \cite{lin2014ipro54, lin2017identifying} randomly extracted non-promoter sequences from middle regions of long coding sequences and convergent intergenic regions in E.coli K-12 genome, which are also 81 bp long. We include these sequences in the non-promoter class.
We have ensured redundancy reduction (no two samples of same class with pairwise sequence identity $\geq$ 0.8) using CD-HIT software (\cite{li2006cd}) on our dataset following \cite{liu2017ipromoter} and \cite{zhang2019multiply}.
We use some recently included promoter samples (experimentally verified) from RegulonDB version 10.7 (\cite{santos2019regulondb}) as our independent test dataset. Our benchmark dataset and test dataset are disjoint. The sample numbers of promoter, its sub types and non-promoter used in 5-fold cross-validation and in independent test purpose have been provided in Table \ref{table:dataset}. 

\begin{table}[]
\begin{tabular}{|c|c|c|}
\hline
\textbf{Classes} & \textbf{\begin{tabular}[c]{@{}c@{}}Benchmark \\ Dataset\end{tabular}} & \textbf{\begin{tabular}[c]{@{}c@{}}Independent\\ Test Dataset\end{tabular}} \\ \hline
Promoter         & 2860                                                                            & 256                                                                          \\ \hline
Non-Promoter     & 2860                                                                            & 0                                                                           \\ \hline
$\sigma^{24}$-promoter               & 484                                                                             & 30                                                                           \\ \hline
$\sigma^{28}$-promoter               & 134                                                                             & 4                                                                           \\ \hline
$\sigma^{32}$-promoter               & 291                                                                             & 13                                                                           \\ \hline
$\sigma^{38}$-promoter               & 163                                                                             & 10                                                                           \\ \hline
$\sigma^{54}$-promoter               & 94                                                                              & 0                                                                           \\ \hline
$\sigma^{70}$-promoter               & 1694                                                                            & 199                                                                          \\ \hline
\end{tabular}
\caption{Class-wise sample numbers in datasets used}
\label{table:dataset}
\end{table}

\subsection{Mathematical Formulation of DNA Sequence} \label{feature}
The goal of formulating an effective mathematical expression which represents a nucleotide sequence is feature extraction. It is challenging to find an appropriate way of expressing a biological sequence such that sufficient sequence-order information is kept. Computational methods require vector representation for prediction or classification tasks (\cite{chou2015impacts}). We consider vector representation of two categories for our work. We describe them in the following subsections.

\subsubsection{Original Nucleotide Sequence}
A DNA sequence can be expressed as follows:\\
$D = N_1, N_2, N_3, \cdots, N_L$\\
where, $L$ is the length of the DNA sequence and $N_i \in \{A, T, C, G\}$.
Although a DNA sequence comprising of nucleotides do not show any distinguishing property when looked at visually, deep learning based models are powerful enough to infer various distinguishing features from local patterns if we can represent such sequence with appropriate mathematical representation (\cite{umarov2017recognition}, \cite{singh2016predicting}, \cite{xu2016sd}). In a DNA sequence, there can be four types of monomers such as - A, T, C and G. So, our monomer representation of each DNA sample is a $81 \times 4$ size two dimensional matrix (each sequence is 81 nucleotide long in our dataset). Each nucleotide is represented by a one hot vector (1 in one position, all other positions 0) of size four. 

We also construct an overlapping trimer representation of each DNA sequence. Each codon corresponding to a single amino acid is a combination of three nucleotides. Full set of codons form the genetic code. This is why trimers have special significance. In the $L$ length DNA sequence mentioned in this subsection, there are total $L-2$ overlapping trimers which are as follows:\\
$N_1N_2N_3, N_2N_3N_4, N_3N_4N_5, \cdots, N_{L-2}N_{L-1}N_L$\\
There can be $4^{3} = 64$ kinds of possible trimers. We represent each trimer with a one hot vector of size 64. Thus, each DNA sample is represented by a $79 \times 64$ size two dimensional matrix.    

\subsubsection{Structural Properties}
Structural property refers to specific characteristics of DNA molecule  such as stability, rigidity or curvature (\cite{meysman2012dna}). Conformational properties are related to static DNA structure (geometrical property) while physicochemical properties are related to dynamic DNA structure (potential to change in conformation). These properties play an important role in promoter prediction and classification (\cite{abeel2008generic}, \cite{bansal2014role}). \cite{chen2014pseknc} constructed PseKNC-General tool which can convert DNA sequence dataset into pseudo nucleotide compositions providing many choices of physicochemical combinations. This tool provides 90 physicochemical properties (role, twist, tilt etc) for each of the 16 possible dimers and 12 physicochemical properties (trinucleotide GC content, consensus role, consensus rigid etc) for each of the 64 possible trimers. We implement physicochemical property wise normalization (subtract mean and divide by standard deviation) so that each property gets equal chance to act as distinguishing property.     
In $L$ length DNA sample, there are $L-1$ overlapping dimers such as
$N_1N_2, N_2N_3, N_3N_4, \cdots, N_{L-1}N_L$. We replace each of these dimers with the 90 physicochemical properties and get a $80 \times 90$ size two dimensional matrix for each 81 length DNA sequence sample. Similarly, there are total $L-2$ overlapping trimers such as $N_1N_2N_3, N_2N_3N_4, N_3N_4N_5, \cdots, N_{L-2}N_{L-1}N_L$. We replace each of these trimers with the 12 physicochemical properties and get a $79 \times 12$ size two dimensional matrix for each 81 length DNA sequence sample.

\subsection{Model Architecture}
\begin{figure}[h]
    \centering
    \includegraphics[width=1.0\columnwidth]{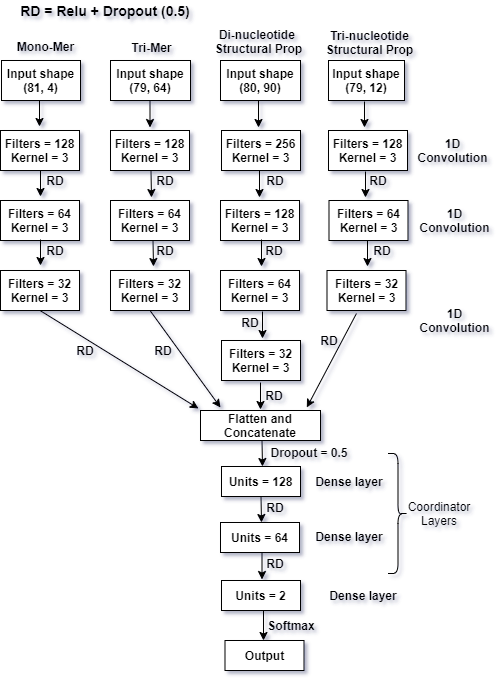}
    \caption{Model Architecture}
\label{fig:model_architecture}
\end{figure}

We use four kinds of feature representations for our model - monomer sequence matrix, trimer sequence matrix, dimer physicochemical property matrix and trimer physicochemical property matrix (two dimensional) of dimension (81, 4), (79, 64), (80, 90) and (79, 12), respectively as described in Subsection \ref{feature}. We provide our model architecture in Figure \ref{fig:model_architecture}. Each of these four unique representations of a sample sequence is passed through a separate one dimensional convolutional neural network (1D CNN) branch parallely as shown in the figure. 

1D CNN has shown its potential and significance in recent studies (\cite{chen2017improving}, \cite{zhou2015icrc}, \cite{oh2018learning}) related to local feature extraction and sequence data classification when the positions of the existing local features are not important. Each of the four branches of 1D CNN works as automatic distinguishing feature extractors for our classification task. The leftmost branch of Figure \ref{fig:model_architecture} gets us significant local combinations of single nucleotides as distinguishing features. The second branch learns locally important combination of codons. The third branch provides us with structurally significant dimer combinations using physicochemical properties. The fourth branch performs exactly the same task but only this time on trimer combinations.

Each of the four different branches learns important distinguishing features from local sequence patterns. In order to perform a successful classification, we need to have a way to combine these independently learnt and extracted features. Each of these four branches return a matrix (two dimensional) of dimension $m_i \times n_i$, where the branch number is $i$. We flatten each of these matrices into a one dimensional vector and concatenate all four of them. The resultant one dimensional concatenated vector is of length $\sum_{i=1}^{4} m_i \times n_i$. This now works as a feature vector. Instead of using this feature vector directly for classification, we pass this vector through densely connected neural network layers so that our model is able to learn successfully the importance of each feature and how to combine them for the classification task at hand. These two dense layers shown in Figure \ref{fig:model_architecture} coming just before the Softmax output dense layer are the coordinator layers.  

We use \textbf{relu} as activation function for each intermediate layer as it has become very popular for its simplicity and effectiveness (\cite{li2017convergence}, \cite{yarotsky2017error}, \cite{agarap2018deep}). In the final layer, we use Softmax activation function for binary classification purpose with two nodes constituting the last dense layer. Neural networks when trained on relatively small amount of data (similar to our case) has a good chance of memorizing the training data instead of learning distinguishing features. Dropout is a regularization method where some layer outputs are ignored in a random manner. Such treatment changes the connectivity of a layer with its previous layer on each epoch of training forcing the model architecture to look different every time (\cite{srivastava2014dropout}, \cite{srivastava2013improving}, \cite{baldi2013understanding}). We use a high dropout rate of 0.5 (Output from a particular layer node is ignored with 50\% probability) after each of our layers (except for the input layer and output layer) to prevent overfitting. 

We use total six binary classifiers for sigma promoter identification and classification into six classes. The details related to the six classifiers have been discusses in Subsection \ref{evaluation}. Each of these classifiers have the same architecture as shown in Figure \ref{fig:model_architecture}. But each of them have different convolution filter and dense layer weights. We provide the .h5 files of each of the six trained models as part of the supplementary information.  

\subsection{Model Selection and Performance Evaluation} \label{evaluation}
Although promoter classification is a multi-class classification problem, the dataset that we use has severe class imbalance problem. For example, there are 1694 samples in $\sigma^{70}$-promoter, the largest promoter subset while only 94 samples belong to the smallest promoter subset $\sigma^{54}$-promoter. Our deep learning based model showed poor training performance when we used the smart undersampling technique introduced for training bTSSfinder tool by \cite{shahmuradov2017btssfinder}. The probable reason is that the minority class $\sigma^{54}$ has only 94 samples and such low number of samples from each class is not enough to train deep learning based models. We have also tried out the popular Synthetic Minority Oversampling Technique (SMOTe) in order to class balance our dataset in accordance with the sample number of our majority class $\sigma^{70}$. Although our model was able to achieve high training performance, performance on validation set was poor which indicates overfitting. The probable cause is that oversampling techniques such as SMOTe fail to produce realistic samples when it comes to high dimensional data (\cite{lusa2013smote}).

To tackle this problem, we have used stage by stage binary classification as shown in Table \ref{table: classification}. The first binary classifier distinguishes between promoter and non-promoter. Each class contains 2860 samples. This number is larger than the largest promoter subset sample number. If the DNA sequence is a promoter, the second binary classifier classifies $\sigma^{70}$-promoter and non $\sigma^{70}$-promoter ($\sigma^{24}$, $\sigma^{28}$, $\sigma^{32}$, $\sigma^{38}$, $\sigma^{54}$). The next largest promoter subset belongs to $\sigma^{24}$-promoter. If the promoter is not $\sigma^{70}$, then the third classifier classifies between $\sigma^{24}$-promoter and non $\sigma^{24}$-promoter ($\sigma^{28}$, $\sigma^{32}$, $\sigma^{38}$, $\sigma^{54}$). This process goes on until we reach a point while we only have two promoter subsets left - $\sigma^{28}$ and $\sigma^{54}$. The last binary classifier distinguishes between these two classes, where $\sigma^{54}$-promoter is the smallest of promoter subsets.

All these six binary classifiers have the same architecture as of Figure \ref{fig:model_architecture}. The only difference is the weights assigned to different network layers because of the difference in training data. For example, the first classifier is trained with promoter vs non-promoter training samples while the last classifier is trained with $\sigma^{28}$ vs $\sigma^{54}$ training samples. The optimizer that we have used to update the weight is \textbf{Adam} (Adaptive moment estimation) while as loss function, we have used \textbf{Categorical Crossentropy}.

The hyperparameters to be tuned in our model architecture are of three types.
\begin{itemize}
    \item \textbf{Layer: } number of convolution filters in each convolution layer, convolution filter size, number of dense layers, number of nodes in each dense layer
    \item \textbf{Function: }choice of activation function in different layers, optimizer and loss function
    \item \textbf{Rate: }learning rate, dropout rate
\end{itemize}
We use 5-fold cross-validation in order to tune our hyperparameters such that we get the best validation performance. The final selected hyperparameter values have been shown in Figure \ref{fig:model_architecture}. We have also experimented with our model after the inclusion of 4-mers as another input branch. Such inclusion did not cause any improvement in result although it caused computational overhead. So, we have not included any branch for 4-mers. We have used an extra layer of filter and kernel for di-nucleotide structural property. This particular branch input has 90 columns which is large compared to the column number of the other three input branches. The extra layer assists the learning of such large number of features. It is interesting to note that for all six of our binary classifiers, this particular architecture shows the best performance. The reason may lie in the fact that all binary classifiers deal with classification related to E. coli sigma promoters.

We have used accuracy (acc), sensitivity (Sn), Specificity (Sp) and Mathew's Correlation coefficient (MCC) as metrics for performance evaluation and comparison with other methods. The metrics are described as follows: \\ \\
$Acc = \dfrac{TP + TN}{TP + TN + FP + FN}$\\ \\
$Sn = \dfrac{TP}{TP + FN}$ \\ \\
$Sp = \dfrac{TN}{TN + FP}$ \\ \\
$MCC = \dfrac{TP \times TN - FP \times FN}{\sqrt{(TP + FP)(TP + FN)(TN + FP)(TN + FN)}}$ \\ \\
All symbols are directed towards binary classification. Here, TP, TN, FP and FN  denote the number of true positive, true negative, false positive and false negative samples depending on model predicted label. Sn and Sp are also known as true positive rate and true negative rate, respectively. These four metrics are widely used for assessing the performance of works related to genome analysis (\cite{chen2014itis}, \cite{chen2015irna}, \cite{ding2014ictx}, \cite{jia2016psuc}, \cite{jia2016psumo}, \cite{qiu2016iptm}, \cite{liu2017ipromoter}, \cite{zhang2019multiply}). Values related to accuracy, sensitivity and specificity lie in the range [0, 1] while for MCC score, the range is [−1, +1]. Higher value indicates better classification ability. 

\begin{table}[]
\begin{tabular}{|c|c|c|c|c|c|}
\hline
\textbf{\begin{tabular}[c]{@{}c@{}}Binary\\ Classifier\end{tabular}} & \textbf{\begin{tabular}[c]{@{}c@{}}Positive\\ Class\end{tabular}} & \textbf{\begin{tabular}[c]{@{}c@{}}Positive \\ Class\\ Sample \\ No.\end{tabular}} & \textbf{\begin{tabular}[c]{@{}c@{}}Negative \\ Class\end{tabular}} & \textbf{\begin{tabular}[c]{@{}c@{}}Negative\\ Class\\ Sample\\ No.\end{tabular}} & \textbf{\begin{tabular}[c]{@{}c@{}}Total\\ Sample\\ No.\end{tabular}} \\ \hline
Model 1                                                              & Promoter                                                          & 2860                                                                               & Non-promoter                                                       & 2860                                                                             & 5720                                                                  \\ \hline
Model 2                                                              & $\sigma^{70}$                                                               & 1694                                                                               & \begin{tabular}[c]{@{}c@{}}$\sigma^{24}$, $\sigma^{28}$, $\sigma^{32}$,\\ $\sigma^{38}$, $\sigma^{54}$\end{tabular}  & 1166                                                                             & 2860                                                                  \\ \hline
Model 3                                                              & $\sigma^{24}$                                                               & 484                                                                                & \begin{tabular}[c]{@{}c@{}}$\sigma^{28}$, $\sigma^{32}$, $\sigma^{38}$,\\  $\sigma^{54}$\end{tabular}       & 682                                                                              & 1166                                                                  \\ \hline
Model 4                                                              & $\sigma^{32}$                                                               & 291                                                                                & $\sigma^{28}$, $\sigma^{38}$, $\sigma^{54}$                                                      & 391                                                                              & 682                                                                   \\ \hline
Model 5                                                              & $\sigma^{38}$                                                               & 163                                                                                & $\sigma^{28}$, $\sigma^{54}$                                                           & 228                                                                              & 391                                                                   \\ \hline
Model 6                                                              & $\sigma^{28}$                                                               & 134                                                                                & $\sigma^{54}$                                                                & 94                                                                               & 228                                                                   \\ \hline
\end{tabular}
\caption{Multi Stage Classification Using Multiple Binary Classifiers}
\label{table: classification}
\end{table}

\subsection{Potential Motif Identification}
\begin{figure}[h]
    \centering
    \includegraphics[width=1.0\columnwidth]{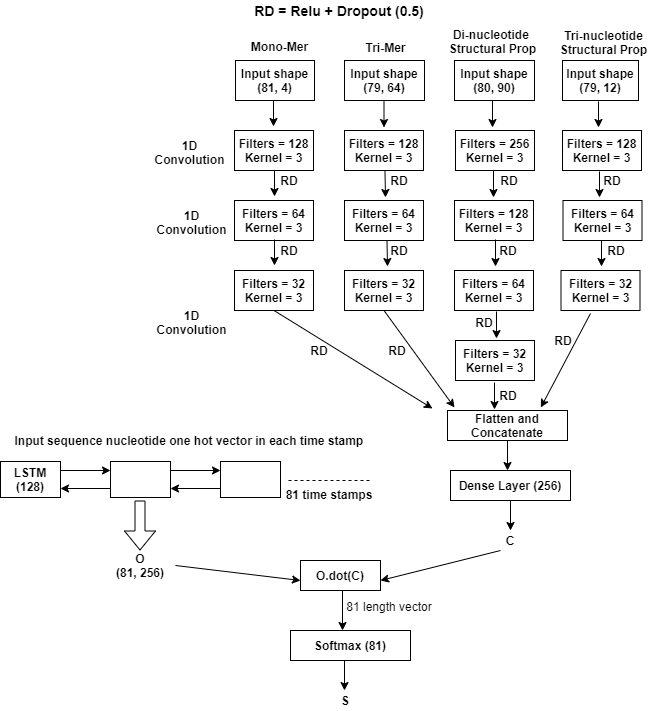}
    \caption{Attention Mechanism Model Diagram}
\label{fig:attention}
\end{figure}
Specific local patterns frequently found in the samples of a class of nucleotide sequences are often important for identifying that particular class. Deep learning based classification models actively search for such potential motifs. We have constructed an attention based model shown in Figure \ref{fig:attention} merging our branched CNN architecture with LSTM based attention mechanism on the input sequence. The 81 size output vector of the \textbf{Softmax (81)} layer of this trained model denote the activation of the corresponding indexed nucleotide of the input sequence. A value greater than $\frac{1}{81}$ denotes activation of the corresponding nucleotide. Using this model, we have identified potential motifs for identification of promoter, $\sigma{28}$, $\sigma{38}$ and $\sigma{70}$ class shown in Table \ref{table: motif}. For example, motif \textbf{AAAAAA} can be found to be activated in 15\% of the available promoter sequences of the benchmark dataset.          

\begin{table}[]
\begin{tabular}{|c|c|c|}
\hline
\textbf{\begin{tabular}[c]{@{}c@{}}Identified\\ Class\end{tabular}} & \textbf{\begin{tabular}[c]{@{}c@{}}Motif\\ Sequence\end{tabular}} & \textbf{\begin{tabular}[c]{@{}c@{}}Active \\ Occurrence\\ Percentage\end{tabular}} \\ \hline
                                                                    & {\color[HTML]{000000} AAAAAA}                                     & 15                                                                                 \\ \cline{2-3} 
                                                                    & {\color[HTML]{000000} ATAAA}                                      & {\color[HTML]{000000} 38}                                                          \\ \cline{2-3} 
\multirow{-3}{*}{\textbf{Promoter}}                                 & {\color[HTML]{000000} AAAAT}                                      & 30                                                                                 \\ \hline
                                                                    & {\color[HTML]{000000} AAAAA}                                      & 25                                                                                 \\ \cline{2-3} 
                                                                    & {\color[HTML]{000000} ATAAA}                                      & 18                                                                                 \\ \cline{2-3} 
\multirow{-3}{*}{\textbf{Sigma28}}                                  & {\color[HTML]{000000} TTAAA}                                      & 13                                                                                 \\ \hline
\textbf{Sigma38}                                                    & {\color[HTML]{000000} CCGCT}                                      & 10                                                                                 \\ \hline
                                                                    & {\color[HTML]{000000} ATATT}                                      & 19                                                                                 \\ \cline{2-3} 
                                                                    & {\color[HTML]{000000} AATAT}                                      & 13                                                                                 \\ \cline{2-3} 
\multirow{-3}{*}{\textbf{Sigma70}}                                  & {\color[HTML]{000000} ATTTT}                                      & 11                                                                                 \\ \hline
\end{tabular}
\caption{Identified potential motifs using attention mechanism}
\label{table: motif}
\end{table}

\subsection{Promoter Prediction in E. coli Genome Sequence}

 \begin{figure}[h]
    \centering
    \includegraphics[width=1.0\columnwidth]{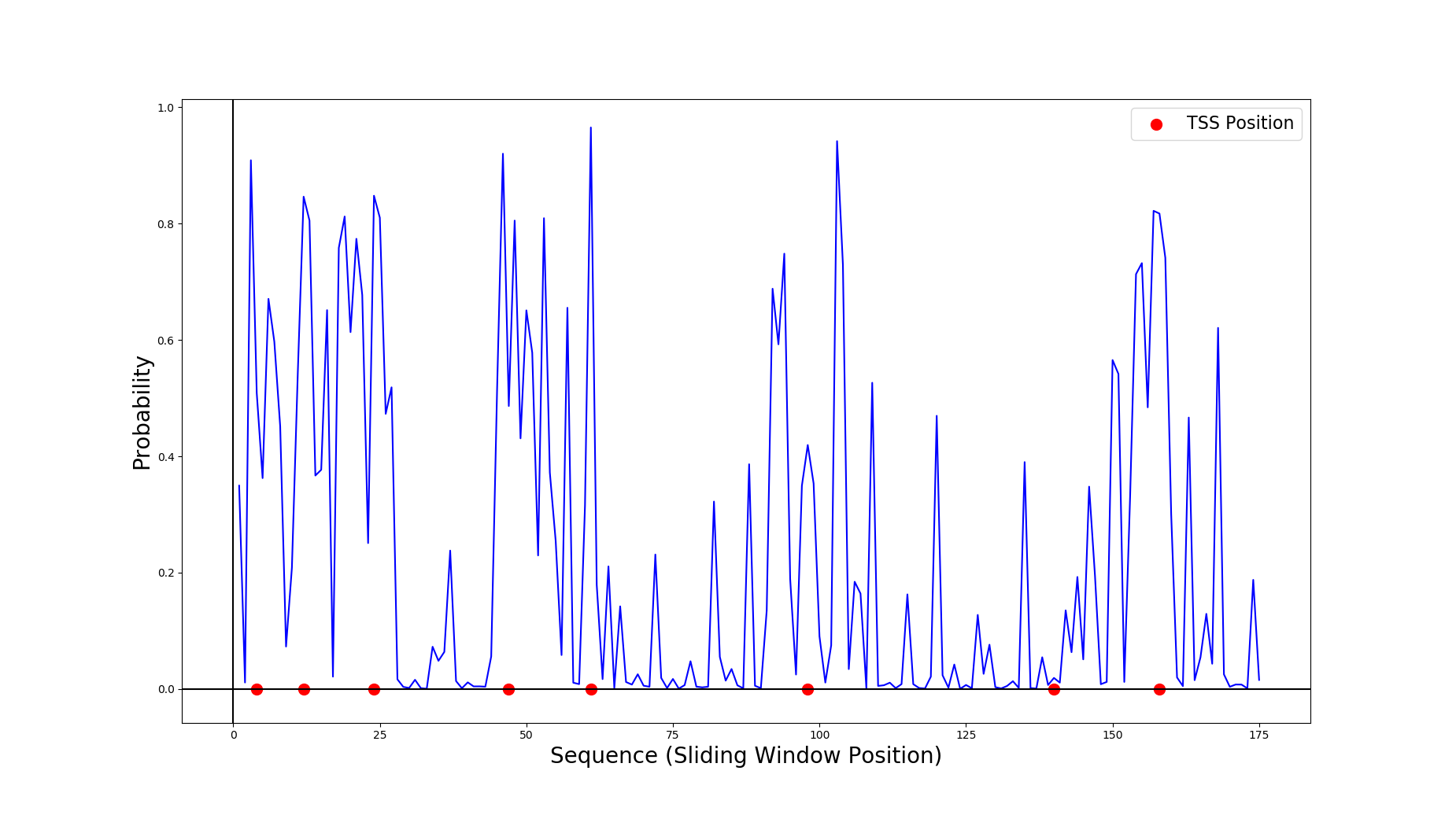}
    \caption{Result of promoter classification model implementation on E. coli genome
    sequence containing 12 genes and 8 TSSs. The large spikes of the graph denote probable promoter regions and the red dots denote given TSS sites}
\label{fig:TSS}
\end{figure}

\cite{umarov2019promoter} used sequence-based deep learning models for identifying TSS regions in long human genome. Although our work is based on promoter classification on 81 length nucleotide sequences, we have tested our model on E. coli genome segment containing 14213 nucleotides and 12 genes obtained from RegulonDB version 10.7 (\cite{santos2019regulondb}). We use sliding window approach on the long genome where window size and stride are both 81. We make promoter prediction on each 81 length window position. The high spikes in Figure \ref{fig:TSS} denote high probability of being promoter. The locations of the TSSs show that our predictor model demonstrates moderate performance in identifying promoter regions in long E. coli genome.

\section{Results and Discussion}
PCSF (\cite{li2006recognition}), vw Z-curve (\cite{song2011recognition}), Stability (\cite{e2014dna}) and iPro54 (\cite{lin2014ipro54}) are some of the state-of-art tools which can identify E. coli sigma promoters. But they do not have the ability of sigma promoter classification. The only two tools with promoter classification capability are iPromoter-2L (\cite{liu2017ipromoter}) and MULTiPly (\cite{zhang2019multiply}).

In order to compare our proposed method with the state-of-the-art promoter identification and classification tools, a consistent benchmark dataset and similar validation methods are required. So, we have used the same training dataset and 5 fold cross-validation used by MULTiPly (\cite{zhang2019multiply}) and iPromoter-2L (\cite{liu2017ipromoter}). Performance comparison between the methods used for promoter identification has been shown in Table \ref{table: performance_promoter}.
The superior performance of proposed iPromoter-BnCNN tool can be seen in all four performance metrics for this particular task.

We demonstrate performance comparison between MULTiPly, iPromoter-BnCNN and iPromoter-2L in Table \ref{table: performance} on sigma promoter classification. Tool iPromoter-BnCNN shows superior performance over MULTiPly for all the classification tasks. The sensitivity and specificity of iPromoter-BnCNN for promoter identification and classification are not only higher than MULTiPly but also the values show more consistency. As a result, iPromoter-BnCNN shows considerably higher MCC score than MULTiPly in all cases. Although iPromoter-2L achieved impressive accuracy in $\sigma{32}$ and $\sigma{38}$ classification, there is a large imbalance in Sn and Sp score which indicates class bias. The MCC score of this tool is much lower compared to MULTiPly and iPromoter-BnCNN. It is to note that all the tool results shown in Table \ref{table: performance_promoter} and Table \ref{table: performance} except for iPromoter-BnCNN tool have been obtained from \cite{zhang2019multiply} and \cite{liu2017ipromoter} articles.

We also show comparison of these three tools discussed above on an independent test dataset obtained from RegulonDB version 10.7 (\cite{santos2019regulondb}). The number of class based true positive and false positive results have been provided in Table \ref{table: independent_performance}. Except for $\sigma{70}$ classification, iPromoter-BnCNN shows state-of-the-art performance on independent test dataset as well.
All independent test results have been obtained through running the tools on the independent test dataset.

\begin{table}[]
\begin{tabular}{|c|c|c|c|c|}
\hline
\textbf{Method}                            & \textbf{Sn}     & \textbf{Sp}     & \textbf{Acc}    & \textbf{MCC}  \\ \hline
{\color[HTML]{242021} \textbf{PCSF}}       & 78.9\%          & 70.7\%          & 74.8\%          & .498          \\ \hline
{\color[HTML]{242021} \textbf{vw Z-curve}} & 77.8\%          & 82.8\%          & 80.3\%         & .61           \\ \hline
{\color[HTML]{242021} \textbf{Stability}}  & 76.6\%          & 79.5\%          & 78.0\%          & .562          \\ \hline
{\color[HTML]{242021} \textbf{iPro54}}     & 77.8\%          & 83.2\%          & 80.5            & .61           \\ \hline
\textbf{iPromoter-2L}                      & 79.2\%          & 84.2\%          & 81.7\%          & .634          \\ \hline
\textbf{MULTiPly}                          & 87.3\%          & 86.6\%          & 86.9\%          & .739          \\ \hline
\textbf{iPromoter-BnCNN}                   & \textbf{88.3\%} & \textbf{88.0\%} & \textbf{88.2\%} & \textbf{.763} \\ \hline
\end{tabular}
\caption{Promoter identification performance comparison using 5-fold cross-validation on benchmark dataset}
\label{table: performance_promoter}
\end{table}

\begin{table*}[]
\begin{tabular}{|c|c|c|l|c|c|l|c|c|l|c|c|l|c|c|l|}
\hline
\multirow{2}{*}{\textit{\textbf{\begin{tabular}[c]{@{}c@{}}Performance\\ Metrics\end{tabular}}}} & \multicolumn{3}{c|}{\textbf{sigma24}}                            & \multicolumn{3}{c|}{\textbf{sigma28}}                            & \multicolumn{3}{c|}{\textbf{sigma32}}                            & \multicolumn{3}{c|}{\textbf{sigma38}}                            & \multicolumn{3}{c|}{\textbf{sigma70}}                            \\ \cline{2-16} 
                                                                                                 & \textbf{MU} & \textbf{Bn}     & \multicolumn{1}{c|}{\textbf{2L}} & \textbf{MU} & \textbf{Bn}     & \multicolumn{1}{c|}{\textbf{2L}} & \textbf{MU} & \textbf{Bn}     & \multicolumn{1}{c|}{\textbf{2L}} & \textbf{MU} & \textbf{Bn}     & \multicolumn{1}{c|}{\textbf{2L}} & \textbf{MU} & \textbf{Bn}     & \multicolumn{1}{c|}{\textbf{2L}} \\ \hline
\textbf{Acc}                                                                                     & 91.2\%      & \textbf{93.8\%} & 94.5\%                           & 95.9\%      & 96.1\%          & \textbf{96.8\%}                  & 85.7\%      & 90.6\%          & \textbf{94.4\%}                  & 85.3\%      & 91.6\%          & \textbf{94.7\%}                  & 84.9\%      & \textbf{87.3\%} & 80.7\%                           \\ \hline
\textbf{Sn}                                                                                      & 88.8\%      & \textbf{93.3\%} & 72.5\%                           & 95.9\%      & \textbf{97.8\%} & 42.5\%                           & 82.2\%      & \textbf{91.7\%} & 52.6\%                           & 83.3\%      & \textbf{94.9\%} & 15.3\%                           & 90.4\%      & 91\%            & \textbf{95.3\%}                  \\ \hline
\textbf{Sp}                                                                                      & 92.9\%      & 94.1\%          & \textbf{96.9\%}                  & 91.3\%      & 93.6\%          & \textbf{99.5\%}                  & 88.4\%      & 89.8\%          & \textbf{99.1\%}                  & 86.7\%      & 89.3\%          & \textbf{99.5\%}                  & 76.9\%      & \textbf{82.2\%} & 59.4\%                           \\ \hline
\textbf{MCC}                                                                                     & .818        & \textbf{.873}   & .734                             & .876        & \textbf{.918}   & .571                             & .708        & \textbf{.9\%}   & .652                             & .699        & \textbf{.833}   & .296                             & .685        & \textbf{.737}   & .606                             \\ \hline
\end{tabular}
\caption{Sigma promoter classification performance comparison between MULTiPly (MU),  iPromoter-BnCNN (Bn) and iPromoter-2L (2L) using 5-fold cross-validation on benchmark dataset}
\label{table: performance}
\end{table*}


\begin{table*}[]
\centering
\begin{tabular}{|c|l|l|l|c|c|l|c|c|l|c|c|l|c|c|l|c|c|l|}
\hline
\multirow{2}{*}{\textit{\textbf{Result}}} & \multicolumn{3}{c|}{\textbf{Promoter}}   & \multicolumn{3}{c|}{\textbf{sigma24}}                        & \multicolumn{3}{c|}{\textbf{sigma28}}                        & \multicolumn{3}{c|}{\textbf{sigma32}}                        & \multicolumn{3}{c|}{\textbf{sigma38}}                        & \multicolumn{3}{c|}{\textbf{sigma70}}                        \\ \cline{2-19} 
                                          & \textbf{MU} & \textbf{Bn}  & \textbf{2L} & \textbf{MU} & \textbf{Bn} & \multicolumn{1}{c|}{\textbf{2L}} & \textbf{MU} & \textbf{Bn} & \multicolumn{1}{c|}{\textbf{2L}} & \textbf{MU} & \textbf{Bn} & \multicolumn{1}{c|}{\textbf{2L}} & \textbf{MU} & \textbf{Bn} & \multicolumn{1}{c|}{\textbf{2L}} & \textbf{MU} & \textbf{Bn} & \multicolumn{1}{c|}{\textbf{2L}} \\ \hline
\textbf{TP}                               & 238         & \textbf{245} & 238         & 19          & \textbf{28} & 18                               & 0           & \textbf{1}  & \textbf{1}                       & 5           & \textbf{10} & \textbf{10}                      & \textbf{4}  & 3           & 2                                & 180         & 179         & \textbf{187}                     \\ \hline
\textbf{FN}                               & 18          & 11           & 18          & 11          & 2           & 12                               & 4           & 3           & 3                                & 8           & 3           & 3                                & 6           & \textbf{7}  & 8                                & 19          & 20          & 12                               \\ \hline
\end{tabular}
\caption{Performance comparison between MULTiPly (MU), iPromoter-BnCNN (Bn) and iPromoter-2L (2L) for identifying promoters and their types on the independent test dataset}
\label{table: independent_performance}
\end{table*}

\section{Conclusion}
We have developed iPromoter-BnCNN in this research for sigma promoter identification and classification in E. coli bacteria. Our architecture combines four different kinds of features from each sample through the use of four one dimensional convolution branches along with coordinator dense layers at the end. Our proposed tool recognizes the specific promoter types in a stage by stage manner with the goal of handling the class imbalance problem. Extensive experiments using 5-fold cross-validation on benchmark dataset and performance on independent test set prove the effectiveness of our proposed method. We expect iPromoter-BnCNN to act as a useful automation tool in the world of computational biology. Constructing a species independent promoter identification and classification model is a possible direction towards future research.

\bibliographystyle{natbib}
\bibliography{main}









\end{document}